# Avoiding Thread Stalls and Switches in Key-Value Stores: New Latch-Free Techniques and More

David Lomet
Redmond, WA USA
dlomet@msn.com

Rui Wang
Microsoft Corp
Redmond, WA USA
wanrui@microsoft.com

## ABSTRACT

A significant impediment to high performance in key-value stores is the high cost of thread switching or stalls. While there are many sources for this, a major one is the contention for resources. And this cost increases with load as conflicting operations more frequently try to access data concurrently. Traditional latch-based approaches usually handle these situations by blocking one or more contending threads. Latch-free techniques can avoid this behavior. But the payoff may be limited if latch-free techniques require executing wasted work. In this paper, we show how latch-free techniques exploit delta record updating and can significantly reduce wasted work by using notices, a new latch-free approach. This paper explains how notices work and can solve B-tree index maintenance problems, while avoiding thread switches or stalls. Other opportunities for avoiding thread switches or stalls are also discussed.

## 1 Introduction

### 1.1 Overview

We want to avoid thread stalls and thread switching in our database kernel as these have an adverse impact on performance. By using delta updating on cached pages, where we prepend a delta (an updated record) to prior page state, we can ensure that a thread switch is not required to update a page in cache. When the prepending is done with a compare and swap (CAS) instruction, this becomes a latch-free update that can be executed concurrently with reads and updates to the same page. We pursued this at Microsoft, with the Bw-tree [1] used in SQL Server's Hekaton [2] main memory database and with Cosmos DB [9] where it was used to index documents in real time, i.e. as they are added. Combined with an appropriate latch-free infrastructure, both reads and updates do not need to wait while a page is undergoing concurrent operations.

Using latches to protect page state while updating blocks both concurrent updates and concurrent reads. If these blocks are of short duration, little is lost as a short latch spin is of minor consequence. But under heavier load, latches blocking both reads and updates have a serious impact on performance. In these cases, queues of waiting threads are frequently employed. Here the thread yields the underlying processor to another job. Thread switching has both extra instructions and loss of caching locality. Latch-free techniques, by retaining control in a single thread, reduce code path length and improve processor efficiency by keeping execution on the existing, frequently in-cache, instruction path.

Delta updating can also avoid an I/O even when a page is not in the page cache, as an update (blind update) does not require the presence in cache of the prior page state. A delta update can be prepended to a partial page state, even a page state entirely absent from the page cache. That state may, however, be required when reading a record that misses in the cache. It is at that point that a read I/O is needed. Delta updating can delay that read and amortize it over multiple updates. Eventually, a page consisting of multiple partial pages needs to be unified, but that can be done lazily.

Updated pages, from time to time, need to be expelled from the cache and written to secondary (we assume flash) storage. We can reduce write I/Os by batching pages into large buffers using a log structured store strategy [8]. This also reduces the frequency of thread switching and I/O path execution as most page "writes" are to a large, multi-page, I/O buffer. Read I/Os are harder to avoid. If a record is not present in cache, its page must be read from storage.

### 1.2 Latch-free Difficulties

Using a CAS to prepend a delta update (record) to a cached page is very simple, with high performance and minimal impact on the performance of other operations, even those accessing the same page. However, over time with more and more record deltas updating the page, the usually more common page reads decline in performance as searching a linear list is much slower than searching pages organized to facilitate search, e.g. binary search or in-page indexing [4]. Thus, we will eventually suffer performance degradation for the very common read operation, to avoid thread switches for hot resources, which impede scalability. This is a trade-off as systems require both high performance and scalability.

So how can we reorganize our updates so that the system can maintain high performance along with high scalability using the latch-free approach? We introduce that topic in section 2. Importantly, we introduce NOTICE, which makes reorganization much less costly. Section 3 and section 4 discuss how to apply the NOTICE idea to the usual B-tree forms of structure modification operations (SMOs [7]). In Section 5, we outline, in a more abstract form, how NOTICEs can be used more generically. Subsequent sections describe further system performance techniques and discuss other work. We end with some conclusions.

## 2. Latch-free Concurrency for Page Updating

### 2.1 Latches

Using latches tends to make correct multi-threaded programming easier, by excluding multi-threading in the difficult critical sections where race conditions can corrupt state. For example, if a thread is updating a B-tree node, setting a latch (a Test and Set) to block other threads from accessing the critical section will ensure that no other accessor (reader or writer) will see the node in a transition state. Unfortunately, with the increased number of threads, this latch based critical section approach interferes increasingly with efficient use of threads. Basically, the threads need to either idle (usually a spin wait) until the latch is released, or switch to another task that can be executed. But switching threads is very expensive, both in number of instructions and the impact on processor cache hit ratios. Both strategies are used in varying combinations. All result in interfering with the maximum exploitation of thread computational power. And in high contention work loads, latching seriously limits multi-thread scalability as threads increasingly run into latched nodes.

### 2.2 Latch-free Basics

Latch-free technology, usually using a Compare-and-Swap (CAS) instruction frequently avoids the kind of thread interference faced by latches and has the potential to more fully exploit thread processing capability. Because access to data is not impeded, care must be taken when state changes are made. In the simple case, a new state is created off-line and then installed by switching a pointer from old to new state atomically using a CAS. At least for a while, both new and old state need to be accessible, with old state discarded only after it is no longer accessible by any active thread. This requires the construction of an infrastructure to deal with this, e.g. an epoch mechanism that tracks whether an active thread continues to execute on some part of an older state. This is non-trivial to implement correctly with high performance. But once done [1], it plays only a minor role in the thread execution cost.

### 2.3 Latch-free Delta Updates

The Bw-tree node delta update illustrates how simple and efficient the latch-free approach can be, while also revealing that there are complications. All nodes are accessed via a mapping table, the table index identifying the node and the table entry pointing to the node state. References to nodes are always via the mapping table.

To update a node requires atomically replacing old node with new node. This can be expensive. But for a simple delta update, this is trivial. Offline, we construct an update "delta" that usually contains a record and the operation involving the record that is to be applied to the page. That delta points to the old state. The update occurs when this new delta is installed, using a CAS, that replaces the old state pointer for the node in the mapping table with a pointer to the update delta. As noted, the new update delta then points to the prior node state. This logically changes the node state to include the effect of the update delta. If the CAS succeeds, this delta update induces a state change effective immediately.

A competing update from another thread can, however, sometimes win the race and install its update first. The CAS used on the node mapping table entry to install new state arbitrates which thread wins the race. The loser thread then repeats the process, now using the updated state that resulted from the winning thread's effort. The great advantage of this approach is that the new state shares the old state and minimizes the execution load for the updating thread. Indeed, this is much more efficient than latching the page and then moving records around in the page to accommodate the new update. In addition, no thread is blocked.

### 2.4 Larger Node State Changes

However, there is an added cost for threads accessing a delta updated node. Cost now includes searching a delta list in addition to whatever read-optimized state we started with, called the "base state" or "base node". This cost grows as the delta update list grows.

So periodically, we want to incorporate the list of delta updates into a node's read-optimized base state. Our original effort worked as follows. To the side, we built a new state that consolidated the base state and the effects of the delta updates to create a new optimized base state. Then, with a CAS, we installed a pointer to this consolidated state in the mapping table entry for the node. The logic here is simple. Without latches, we have transformed node state to make accesses to it perform better. But there is a performance penalty. If multiple threads try to install a consolidated state, only one will win the CAS, which is what we want. But the losing threads will also pay the cost of consolidation, which can be nontrivial. We want to avoid this extra cost.

### 2.5. Avoiding Wasted Node Consolidations

We introduce here an idea invented by Rui Wang [10]. The idea is for threads to race to install a delta. The winning thread's delta announces that its thread is responsible for consolidating state. Hence, the CAS race is to install the delta, not to install the consolidated state. This means that the losing threads only lose the work of generating and trying to install this "consolidating" delta. The thread that wins the CAS and installs this delta then, by itself only, builds and installs the consolidated state. We call this delta a notice. We will use notices in other Bw-tree changes as well. A thread consolidating deltas with base state first prepends a cNOTICE to the state of the node. The "c" in "cNOTICE" designates that the purpose of the notice is consolidating the update deltas with base page to produce a new base page. We will use this type of naming convention for our other notices.

Constructing and prepending a cNOTICE is trivial compared with generating the consolidated state. And once posted, it keeps other threads from paying the cost of generating a consolidated state. Only the winner of the race to post the cNOTICE pays the cost of consolidation. Losing threads pay only the cost of trying to post the cNOTICE and failing. Since these other threads are usually threads performing updates, they can continue execution by prepending their updates to the cNOTICE. Only the "winner" of the cNOTICE race pays the cost of node consolidation, on a state that is guaranteed not to be changed by any other thread.

We illustrate the sequence of states involved in a node consolidation in Figure 1. In Figure 1(a), the cNOTICE as been posted, protecting the prior state, which includes here the old base page and some update deltas, from further change. In Figure 1(b) the poster of the cNOTICE has generated a new base page that now includes the previously posted update deltas. While this consolidation is ongoing, other threads may continue to post their update deltas.

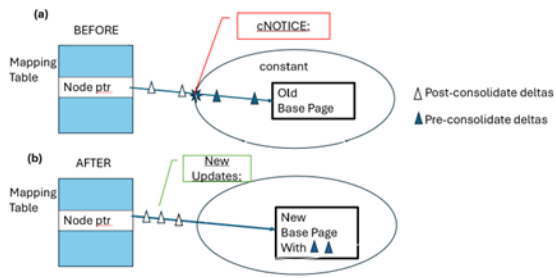

**Figure 1: Using a cNOTICE to consolidate a page. The thread that installs the cNOTICE is the only consolidator, while others continue to read and update the node.**

To ensure that the winner gets its update posted prior to updates of losers, the winner may build a two-delta list, including both the cNOTICE and its update, to prepend to the node state. This permits the cNOTICE to guard node state that includes the winner's update and ultimately consolidates that update into the new base page.

## 2.6 Structure Modification Operations

Larger multi-node changes incur some cost prior to the CAS. These multi-node changes are structure modification operations (SMOs) [7]. Our goal is to minimize this pre-CAS work, since it will be wasted for threads that have lost the CAS. However, we want these CAS loser threads to continue execution smoothly. So competing threads need to prepare a post CAS state so that losing threads can continue their execution productively. It is this post CAS continuation path for losing threads that is lost by loser threads.

However, as with consolidation, the expensive part of the SMO is replacing the state guarded by the notice. And that is done only by the thread winning the CAS.

## 3. Latch-free Node Splits

We used B-link tree style node splits in the original Bw-tree [1]. This permits a node split to be performed in two separate atomic actions, one that splits the node, and another that posts an index term for the new node. For the Bw-tree, splits were done in two atomic actions also, in the same spirit as our prior consolidations. We created the state for a new node with half of the old node entries, copying these entries from old node to new node, and then attempted to install a split delta using a CAS. Multiple threads may attempt the split, meaning CAS losers also performed this new node building work. This is followed with the second B-link atomic action posting an index term at the parent index node for the newly split node using a latch-free node update as described in section 2.

As before, we want to reduce the work done by a thread prior to the CAS by exploiting a notice. In this case we split the page by exploiting a split notice (sNOTICE). The CAS winning thread does most of the node split work. But there is some work that needs to be performed to ensure that CAS losers can proceed, usually with them doing additional updating. We want to minimize that pre-CAS work while ensuring that old node state prior to the sNOTICE is protected from changes other than by the CAS winner. This is illustrated in Figure 2. This work enables loser threads to proceed even before a half-split is completed. We do not include the parent index node update part of the B-link tree split, which is done by a delta update at the parent index node and can be performed after the splitting of the node.

We have one more concern related to storage. We want the new node N to be populated with its appropriate records by the time the split is made persistent and before the old node O shows up in the stable state without the records moved to N. Otherwise, we risk records in O being lost for the stable state should the system crash. Here we describe how Deuteronomy [1] uses its log structured buffer to accomplish this.

1. For new node N, we only make a slot in the mapping table. We allocate space in the I/O buffer used to send data to persistent storage. This space will hold the data states of both N and O. But no data is moved. Allocating this space in the I/O buffer ensures that the split either persists completely or not at all should the system crash. The buffer is not written to storage until the splitter releases its hold on the buffer. Further, O's pre-split state, now split with N, will be in the log structured store buffer prior to further delta updates for O or N arriving.

2. The sNOTICE contains the key we used to define the split and pointers for both N and O mapping table entries, and the location in the I/O buffer of the space allocated for O and N. The sNOTICE is posted without contention at N's mapping table entry. The split defining key becomes the new side link key for O.

3. We use a CAS to install this sNOTICE at O. This situation is shown in Figure 2.  If the CAS fails, we only lose the work in steps 1. and 2. above. No data has been moved, which is the major split execution cost. The sNOTICE is now in place at both O and N, permitting O's data below the sNOTICE to be shared by both. Subsequent accesses to O will encounter the sNOTICE, which directs them to O or N as indicated by the sNOTICE's split key.  Accessors to either N or O execute "normally", except for sharing the state protected by the sNOTICE until the N is fully instantiated and O is cleared of entries that now reside in N.

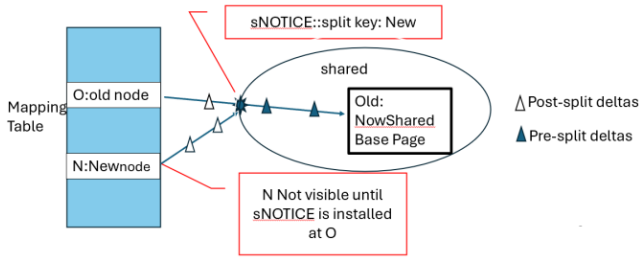

**Figure 2: Node split at step 2 – simplified.  The encircled part is the unsplit node (originally the old node), now guarded by an sNOTICE.  It is read-only, and is now shared between old and new nodes.  Updates can continue at either O or N nodes, depending on their key values.**

4. The sNOTICE CAS winner moves N's data (with keys larger than the sNOTICE split key) from O to N's allocated buffer space. Then the prepended deltas at N are connected to N's state in the buffer, ending the sharing arrangement.  N's side link pointer is set to the side link pointer from O.

5. The records in the shared state less than the split key are now moved to their new space in the buffer.  We now replace the sNOTICE in O with a pointer to the newly formed version of O. The split key in the sNOTICE becomes the side link pointer of O.  We have now completed what has been called a half split, i.e. splitting the data between two nodes.

6. The final step is to update the parent index node with an index term that directs the search from the parent node to the appropriate node below it based on the split key.  This is the second step in a B-link tree split.  If the system crashes prior to accomplishing this, this part of the split can be completed during or after recovery.

## 4. Latch-free Node Merge

Our original Bw-tree paper did not correctly describe the merging of nodes.  The presence of notices makes the merging process easier and more understandable.  And it is even easier here when we restrict slightly the scope of the nodes that can be merged.  We merge only those nodes that are not referenced via the lowest value index term of a parent index node.  Merging data nodes eliminates an index term in the parent index node.  We do not preclude parent index nodes from splitting.  This creates a problem should the split process select to split a parent index node in the region where a node merge is occurring.  Were that to happen, both resultant index nodes of the split might be responsible for accessing part of the key space of the data merged node.  The merge steps are below.

1. A pNOTICE is posted to the parent index node P of the data nodes M (the resulting merged node) and D (the node to be merged into M).  It indicates which key space adjacent index terms reference the data nodes (M and D) that are to be merged, hence identifying also the node to be removed (the higher key order one D). It is posted to P using a CAS to ensure that P cannot be split within the newly merged child node key space.

2. Now the parent no longer logically contains an index term for D, and all searches for its data proceed via the lower key order node M.  M contains a side pointer to D and a search will follow the side pointer should the data in D be read or updated.  This is the opposite of the node split case.  During a split, one node contains data intended for two, while here two nodes contain data intended for one. We then prepend a dNotice at D to prevent prepending of additional updates to D, i.e. to ensure that no updates for D that have slipped through before the pNotice was posted but after dNotice was prepended can be posted to D; instead, these updaters are redirected back to M, where they can post a delta update on M.

3. Finally as shown in Figure3, we post an mNotice at M so that the contents of both M and D can no longer be changed, and their contents can be consolidated. The mNotice is similar to a cNotice except that it guards both M and D.

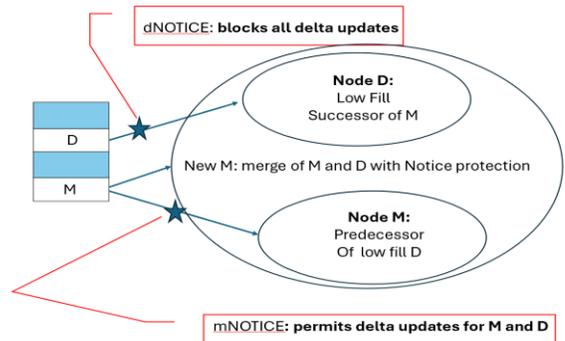

**Figure 3: Node merge at step 3 - simplified.  The large encircled part is the new merged node M containing data from both M and D at the time the mNotice was prepended.  Updates can continue at M, but not at D.  Subsequent updates for D are also prepended at M.**

4. Now we can execute the node merge, which is the "consolidation" of M and D.  This removes D and the dNotice, and replaces the old M with the new M which has merged old M with D. The pNotice can then be removed from P via a consolidation.  And so can the mNotice.

The heavy-duty part of the node merge is consolidating entries from M and D into a single optimized and contiguous storage area.  And, as before, this is done after placing the required notices to protect the state from changing during the merge. Losers of the race to post the pNOTICE at the parent can proceed with their reads or updates, assured that the winning pNOTICE thread will execute the node merge. And any ongoing updates will be either protected by a notice or redirected by it.

## 5. A NOTICE Latch-free Paradigm

There is a common thread (excuse the pun) going through how we use notices to provide non-blocking concurrent read and write access for shared data under contention.

1. Delta updates are essential so that an unchanging pre-delta state can be safely read, and the state can be updated while both reads and updates are concurrently in progress.

2. Notices facilitate light-weight conflict resolution when a thread wishes to make an update that is transformative as opposed to delta based. A notice is prepended in the same way as a delta update. It permits concurrent reads and delta-based updates.

3. Notices guard part of the state whose transformation (update) cannot be accomplished via delta updates. The state to which the notice is prepended cannot be altered by any other thread approaching that state along the access path on which the notice is placed.

4. Notices must be prepended on all paths to the part of the state being transformed. This ensures that the notice protected part of the state cannot be changed except by the notice(s) poster.

5. A thread competing with a notice must ensure that should it win the notice race, the loser threads can continue their operations successfully without needing to change the notice-protected state, e.g. by making a delta update.

6. The notice guarded state being transformed is made read-only. It is transformed by replacement in an atomic action attaching it to the involved states on notice protected paths.

7. Notices can be logically disabled after serving their concurrency purpose by using a "void" marker. They can usually be removed as well, though that takes more work.

8. Latch-free infrastructure needs to be present to perform garbage collection when it is safe to do so. Deuteronomy used an epoch approach [3].

As with other conflict resolution methods, e.g. latches, care must be taken. For example, with latches, deadlocks are possible without proper ordering among the latches. Care must also be taken with notices, their placement and their ordering, to ensure correctness.

## 6. Other Latch-Free Considerations

### 6.1 Representing Lists

Storage management can be expensive. Further, traversing linked lists can involve first accessing the pointer element, and then accessing the data associated with the pointer element. The pointers may be managed separately, both in their allocation and in their garbage collection, as shown in Figure 4(a).

What we suggest here is to prepare a list storage element that is large enough to hold both a list pointer and the associated list element, as in Figure 4(b). This is not a new idea, but it can have a noticeable impact on performance. First, it cuts in half the number of storage allocations per list element. Second, data adjacent to the list pointer (Next in the figure.) is more likely to be in processor cache with the pointer, improving thread execution performance.

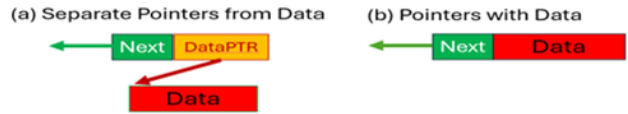

Figure 4: List formatting of two forms. The left form requires 2X the allocations of the right and is more likely to trigger a processor cache fault when "Data" is accessed.

### 6.2 Thread Failures

The "all-at-once" state replacement at the time of the CAS, while more expensive in execution, does have the advantage that should the CAS winner thread die for some reason prior to installing the new state, there is always another thread that will do the work instead. However, we have not described what happens when the thread posting a NOTICE dies after posting the NOTICE but before installing the updated state.

Instead of a notice identifying just a winning thread, the notice can have a time-out whereby the notice will no longer stop another thread from doing the job previously awarded to the notice posting winner. One might use clock time for this purpose, but using epoch numbers might be more effective. A notice posting thread can include in the notice the epoch in which it is executing. After two or three epochs, if the work has not been done, another thread can perform it instead. A CAS prevents duplicate winners from installing changes by altering the notice seen by another CAS.

### 6.3 Updating Pages Not in the Cache

Cosmos DB, originally Document DB [9], wanted to have documents instantly indexed by key word as they were entered into a database of documents, termed as "real-time indexing". If one must read in the pages for all key words of a new document so that these pages can be updated to include the new document, real time indexing will not be possible. Document DB chose to use the Bw-tree to index documents not only for how it performed under most circumstances, but because it supported this real-time indexing.

Each page of the Bw-tree has an entry in a vector-based mapping table. The mapping table references the page either in memory or on secondary storage. The mapping table exploits a log structured store (see 6.4) that enables pages to be relocated. The Bw-tree has this real-time indexing capability because of delta updating. Should a page being updated not be in cache, its mapping table entry will contain its location on flash storage. This permits the page to be delta updated since delta updates do not require access to the prior state of the page. The prior state is linked to the new delta update so that when the page is later read, the full state of the page can be assembled.

### 6.4. Exploiting Log Structured Storage

Contention for resources is not the only source of thread stalls or thread switches. I/O operations, because of flash latency, can produce the need to switch threads. Log structuring is a technique to batch page writes to secondary storage [10]. Deuteronomy uses log structuring to batch writes. Instead of one I/O operation per page, one has one I/O per batch. The pages in the batch do not experience a thread switch when placed in the log buffer. And allocating a page in buffer can be done with a Fetch-and-Increment (FAI) instruction so the operation of the buffer is latch-free and without any thread stall. Only when the buffer is full is more work required, but the Bw-tree makes this process latch-free as well.

This requires work to virtualize secondary storage, but in our experience, it always substantially reduced the number of I/O writes required [8]. Sadly, we know of no way, when dealing with page reads, to similarly reduce I/Os.

### 7. Other Work

So far as we know, the Log Structured Store backed Bw-tree (as the Data Component (DC) of Deuteronomy) and its follow-on Bw$^e$-tree [10], are the only key-value stores exploiting latch-free techniques. Competing storage engines mostly use conventional latching. We compared our DC with other key-value stores. We used Berkley DB [12] in our early benchmarking and its performance was not close to the Bw-tree. RocksDB [13] is a very widely used storage engine, partly because it is open source. Its record updates can be executed without reading data from secondary storage, a valuable property shared by Deuteronomy. In our benchmarking, Deuteronomy performed better [10]. But Deuteronomy is not open source, so it is used in only a few places [2,9,10].

One paper compared the Bw-tree to purely in-memory access methods and, not surprisingly, found the in-memory methods faster [11]. Other main memory methods also demonstrated greater performance, but these papers did not assess cost-performance. We did a cost-performance comparison [5] with MassTree [6], a very fast main memory system, and found MassTree to be faster, but it used much more main memory, leading us to the conclusion that only if you required stellar performance irrespective of cost would you choose it over Deuteronomy and its Bw-tree, which reduces cost by being more memory efficient AND by moving data out of a main memory cache to secondary storage.

Usually hot data, which one wants in the main memory cache, is substantially smaller than cold data, which should be on secondary storage until it is accessed and becomes hot. This is the way that caching-based systems balance cost vs performance, and why they are so widely used. All the large commercial database systems are caching based systems for exactly this reason. Customers want to reduce "COGS" (the cost of goods and services), so long as performance is adequate.

### 8. Conclusions

This paper shows, with a second look, how to greatly improve latch-free technology using notices, avoiding redundant work and making large state changes such as occur in B-tree type access methods simpler and more efficient. This freedom from thread interruptions gives latch-free approaches a substantial performance and scalability edge, provided the rest of the code paths are short. Notices enable short paths by putting expensive code paths after establishing race condition winners.

Reducing thread switching is one of the surest ways to improve performance. A thread switch "trashes" the processor cache and can result in very substantial and often unseen overhead should the switching involve the operating system. Using latch-free techniques does not always have the shortest code paths when contention is ignored. But by providing a path for contention losers to continue with productive execution when they lose during race conditions, avoids the large thread switching cost.

It is possible to extend this general approach to transactional conflicts as well, which we have done in [3] and will explain in an appendix of the full paper.

### Acknowledgements

Our thanks to Justin Levandoski, Sudipta Sengupta and Ryan Stutsman, who participated in the Deuteronomy implementation and contributed several of its innovations.

### REFERENCES

[1] Justin J. Levandoski, David B. Lomet, Sudipta Sengupta: The Bw-Tree: A B-tree for new hardware platforms. *ICDE 2013*: 302-313

[2] Justin J. Levandoski, David B. Lomet, Sudipta Sengupta, Adrian Birka, Cristian Diaconu: Indexing on modern hardware: Hekaton and beyond. *SIGMOD Conference 2014*: 717-720

[3] Justin J. Levandoski, David B. Lomet, Sudipta Sengupta, Ryan Stutsman, Rui Wang: High Performance Transactions in Deuteronomy. *CIDR 2015*

[4] David B. Lomet: The Evolution of Effective B-tree: Page Organization and Techniques: A Personal Account. SIGMOD Rec. 30(3): 64-69 (2001)

[5] David B. Lomet: Cost/performance in modern data stores: how data caching systems succeed. *DaMoN 2018*: 9:1-9:10

[6] Y. Mao, E. Kohler, R. T. Morris. Cache Craftiness for Fast Multicore Key-Value Storage. In EuroSys, 2012, pp. 183-196.

[7] C. Mohan, Don Haderle, Bruce G. Lindsay, Hamid Pirahesh, Peter M. Schwarz: ARIES: A Transaction Recovery Method Supporting Fine-Granularity Locking and Partial Rollbacks Using Write-Ahead Logging. ACM Trans. Database Syst. 17(1): 94-162 (1992)

[8] Mendel Rosenblum, John K. Ousterhout: The Design and Implementation of a Log-Structured File System. SOSP 1991: 1-15…

[9] Dharma Shukla, Shireesh Thota, Karthik Raman, Madhan Gajendran, Ankur Shah, Sergii Ziuzin, Krishnan Sundaram, Miguel Gonzalez Guajardo, Anna Wawrzyniak, Samer Boshra, Renato Ferreira, Mohamed Nassar, Michael Koltachev, Ji Huang, Sudipta Sengupta, Justin J. Levandoski, David B. Lomet: Schema-Agnostic Indexing with Azure DocumentDB. *Proc. VLDB Endow. 8(12)*: 1668-1679 (2015)

[10] Rui Wang, Xinjun Yang, Feifei Li, David B. Lomet, Xin Liu, Panfeng Zhou, Yongxiang Chen, David Zhang, Jingren Zhou, Jiesheng Wu: Bwe-tree: An Evolution of Bw-tree on Fast Storage. *ICDE 2024*: 5266-5279

[11] Ziqi Wang, Andrew Pavlo, Hyeontaek Lim, Viktor Leis, Huanchen Zhang, Michael Kaminsky, David G. Andersen: Building a Bw-Tree Takes More Than Just Buzz Words. *SIGMOD Conference 2018*: 473-488.

[12] Wikipedia: https://en.wikipedia.org›wiki›Berkeley_DB

[13] Wikipedia: https://en.wikipedia.org › wiki › RocksDB